\documentclass[%
 amsmath,amssymb,
preprint,
]{revtex4-1}

\usepackage{graphicx}
\usepackage{dcolumn}
\usepackage{bm}

\usepackage[utf8]{inputenc}
\usepackage[T1]{fontenc}
\usepackage{mathptmx}
\usepackage{etoolbox}

\makeatletter
\def\@email#1#2{%
 \endgroup
 \patchcmd{\titleblock@produce}
  {\frontmatter@RRAPformat}
  {\frontmatter@RRAPformat{\produce@RRAP{*#1\href{mailto:#2}{#2}}}\frontmatter@RRAPformat}
  {}{}
}%
\makeatother

\usepackage{xcolor}

\usepackage{soul} 

\newcommand{\commentout}[1]{}

\newcommand{\SpeciesFlux}{\mathcal{F}}
\newcommand{\ReversibleStress}{\mathcal{R}}
\newcommand{\WhiteNoiseMass}{\mathcal{Z}}
\newcommand{\WhiteNoiseMomentum}{\mathcal{W}}
\newcommand{\ViscousTensor}{\boldsymbol{\tau}}

\begin{document}

\preprint{AIP/123-QED}

\title[Comment]
{Comment on ``Brownian motion of droplets induced by thermal noise''}
\author{J.B. Bell*} \email{jbbell@lbl.gov.}
\author{A. Nonaka}
\affiliation{Lawrence Berkeley National Laboratory}
\author{A.L. Garcia}
\affiliation{Department of Physics \& Astronomy, San Jose State University}%

\date{\today}

\begin{abstract}
We simulate phase separated fluids using the Cahn-Hillard fluctuating hydrodynamic (CH-FHD) model and measure the statistical properties of capillary waves generated by thermal fluctuations.
Our measurements are in good agreement with stochastic lubrication theory and molecular dynamics simulations but differ significantly from recent CH-FHD results by Zhang \textit{et al.} (\textit{Phys. Rev. E} \textbf{109} 024208 (2024)). Specifically, we find that capillary wave statistics at thermodynamic equilibrium are independent of transport properties, namely viscosity and species diffusion.

\end{abstract}

\maketitle

\section{Introduction}


Spontaneous thermal fluctuations continuously excite capillary waves on liquid interfaces.
Driven by the interplay between surface tension and thermal energy, capillary wave theory (CWT) predicts a characteristic spectrum at thermodynamic equilibrium that results from the equipartition of energy. 
Understanding the dynamics and statistical properties of these nanoscale waves plays a crucial role in diverse phenomena, ranging from the stability of thin liquid films to the coalescence of droplets.

Fluctuating hydrodynamics (FHD) is a useful theoretical model for incorporating thermal fluctuations in fluid mechanics.\cite{Landau_59,CapWaveFHD_1983,Zarate_07} 
For sharp interfaces in the lubrication approximation, the FHD equations reduce to a single stochastic PDE for the capillary wave height.\cite{ThinFilmGrun2006}
This stochastic lubrication theory is in agreement with molecular dynamics (MD) simulations in predicting the pinching of a liquid jet~\cite{Nanojets_2000,Eggers_02} and
in describing various thin film phenomena, including rupture~\cite{ThinFilmMecke2001,ThinFilmGrun2006, CapWaveThinFilm_2021,ThinFilmFluct_2021,CapWaveFluct2023}.

Alternative FHD formulations are based on diffuse interface models, such as the Allen-Cahn and the Cahn-Hillard models \cite{CahnHilliard:1958}.
Numerical schemes for the latter model (CH-FHD) have been developed and applied to a variety of problems.\cite{MultiphaseFHD2014,RTIL,Gallo2022} 
A recent paper by Zhang~\textit{et al.}~\cite{Brownian_CA_CH_PRE2024} presented a CH-FHD scheme that showed anomalous results. 
Specifically, they found that the spectrum of capillary wave height varies with viscosity, both in amplitude and in the power law coefficient (see Fig.~1 in \cite{Brownian_CA_CH_PRE2024}). 
In this Comment we present a CH-FHD model for which this spectrum is \emph{independent} of viscosity, in agreement with CWT. 
We also show that our CH-FHD simulation results for interface height fluctuations are in quantitative agreement with theory. 
Finally, we discuss possible sources for the discrepancy between our findings and the anomalous results of Zhang~\textit{et al.}\cite{Brownian_CA_CH_PRE2024}


\section{Cahn-Hillard Fluctuating Hydrodynamics}

The multispecies fluctuating dynamics formulation used here is essentially the same as that described in Barker \textit{et al.}\cite{BrynRP2023}.
Consider a binary mixture of similar species with a specific free energy density ~\cite{CahnHilliard:1958}, 
\begin{align}
\frac{\mathcal{G}}{\rho k_B T} =
c \ln c + (1-c) \ln (1-c) 
+ \chi c (1-c) 
+  \kappa |\nabla c |^2 
\label{eq:CHfree} 
\end{align}
where $\rho$ is the mass density and $c$ is the mass fraction of one of the species.
For interaction coefficient $\chi > 2$ the mixture phase separates into concentrations $c_{e,1}$ and $c_{e,2}$ given by the solution of
\begin{align}
\ln \left( \frac{c_e}{1-c_e}\right) = \chi (2 c_e - 1) 
\end{align}
The surface energy coefficient is $\kappa$ and the surface tension is
\begin{align}
    \gamma = n k_B T \sqrt{2 \chi \kappa} ~\sigma_r(\chi)
    \label{eq:SurfaceTension}
\end{align}
where $n = \rho/m$ is the number density and $\sigma_r \simeq O(1)$.
The characteristic length scale for the interface thickness is $\ell_\mathrm{c} = \sqrt{2\kappa / \chi}$ and for capillary wave fluctuations is $\ell_* = \sqrt{ k_B T / \gamma}$.

The incompressible flow equations for constant $\rho$ are
\begin{align}
\partial_t( \rho c) + \nabla \cdot(\rho u c) =& \nabla \cdot {\SpeciesFlux}   \nonumber \\
\partial_t( \rho u) + \nabla \cdot(\rho u u) + \nabla \pi =& \nabla \cdot {\ViscousTensor}  + \nabla \cdot \ReversibleStress \nonumber \\
\nabla \cdot u =& 0
\label{eq:low_mach_eqs}
\end{align}
where $u$ is the fluid velocity and $\pi$ is a perturbational pressure.
Here, $\SpeciesFlux$, $\ViscousTensor$, and $\ReversibleStress$ are the species flux, viscous stress tensor, and the interfacial reversible stress, respectively.

In fluctuating hydrodynamics the dissipative fluxes are written as the sum of deterministic and stochastic terms. 
The species flux is $\SpeciesFlux = \overline{\SpeciesFlux} + \widetilde{\SpeciesFlux}$ where the deterministic flux is
\begin{align}
\label{eq:DetSpeciesFlux}
\overline{\SpeciesFlux} = \rho D \left ( \nabla c - 2 \chi  c (1-c) \nabla c +  2c(1-c) \kappa \nabla \nabla^2 c \right )
\end{align}
and $D$ is the diffusion coefficient. 
The stochastic flux is
$\widetilde{\SpeciesFlux} = \sqrt{2 \rho m D ~c (1-c)}  ~\WhiteNoiseMass$
where $\WhiteNoiseMass(\mathbf{r},t)$ is a standard Gaussian white noise vector with uncorrelated components.
The viscous incompressible stress tensor is
$\ViscousTensor =  \overline{\ViscousTensor} +  \widetilde{\ViscousTensor}$
where the deterministic component is
$\overline{\ViscousTensor} = \eta [\nabla u + (\nabla u)^\mathrm{T}]$.
The stochastic contribution to the viscous stress tensor is 
$\widetilde\ViscousTensor = \sqrt{\eta k_B T}({\WhiteNoiseMomentum} + {\WhiteNoiseMomentum}^\mathrm{T}),
$ where 
${\WhiteNoiseMomentum}(\mathbf{r},t)$ is a standard Gaussian white noise tensor
with uncorrelated components.
Finally, the interfacial reversible stress is
\begin{align}
\ReversibleStress = n k_B T \kappa \left [\frac{1}{2} |\nabla c|^2 \mathbb{I} - \nabla c \otimes \nabla c \right].
\label{eq:reversible_stress}
\end{align}
Note that since $\ReversibleStress$ is a non-dissipative flux there is no corresponding stochastic flux.

For numerical calculations the system of equations (\ref{eq:low_mach_eqs}) is discretized using a structured-grid finite-volume approach with cell-averaged concentrations and face-averaged (staggered) velocities with standard spatial discretizations. The algorithm uses an explicit discretization of concentration coupled to a semi-implicit discretization of velocity using a predictor-corrector scheme for second-order temporal accuracy.
The discretized Stokes system is solved by a generalized minimal residual (GMRES) method with a multigrid preconditioner, see \cite{cai:2014}.
The explicit treatment of the concentration equation introduces a stability limitation on the time step of
\begin{align}
D \left( \frac{12}{\Delta x^2} + \frac{72 \kappa}{\Delta x^4} \right) \Delta t \leq 1
\label{eq:StableDt}
\end{align}
where $\Delta x$ is the mesh spacing.
The numerical scheme is based on methods introduced in~\cite{donev2014low, Donev_10, RTIL}; details are discussed in Barker \textit{et al.}\cite{BrynRP2023} and its Supporting Information.


\section{Capillary Wave Simulations}

Unless otherwise specified, the physical parameters used in all simulations are as follows: 
mass density, $\rho = 1.0~\mathrm{g/cm}^3$, 
molecular mass, $m = 1.8\times 10^{-22}$~g, 
temperature $T = 300$K. 
The Flory interaction parameter $\chi = 3.78$ so
the equilibrium concentrations are $c_{1,e} = 0.027285$ and $c_{2,e} = 1 - c_{1,e} = 0.972715$.
The surface energy coefficient is $\kappa = 4.0\times 10^{-14}~\mathrm{cm}^{2}$
giving a surface tension, $\gamma = 32.85$~dyne/cm.
For these values $\ell_\mathrm{c} = 1.45$~nm, $\ell_* = 0.355$~nm, and
the interface thickness is roughly 3.6~nm.
We consider a range of values for the shear viscosity and diffusion coefficient. 
Specifically, simulations used $\eta = 10^{-1}, 10^{-2}$ or $ 10^{-3}$~poise with diffusion coefficients of $D = 2 \times 10^{-4}$,  $2 \times 10^{-5}$, or  $2 \times 10^{-6}~\mathrm{cm^2/s}$.

In general, the simulation mesh was a quasi-2D grid with $N_x \times N_y \times 1$ grid points.
The mesh spacing was $\Delta x = \Delta y = 1.0$~nm.
The depth in the $z$ direction, $L_z = \Delta z$, only serves to set the magnitude of the noise.
The time step was $\Delta t = 0.4$~ps for the two smaller values of $D$, which corresponds to approximately 0.25 of the maximum stable time step for the $D=2\times 10^{-5}~\mathrm{cm^2/s}$ (see Eq.~\ref{eq:StableDt}). For the largest value of $D$ we reduced $\Delta t$ by an order of magnitude for stability reasons.

A basic validation test of the code is the measurement of the Laplace pressure $\delta p$ for 2D droplets of various radii $R$.
Simulation data presented in Table~\ref{tab:Laplace} verifies the code captures the correct pressure jump, namely, 
$\delta p = \gamma R$.
Barker \textit{et al.}\cite{BrynRP2023} discuss other validation tests of our CH-FHD code.

\begin{table}[]
    \centering
    \begin{tabular}{l||c|c|c||c|c|c}
        Resolution~    &  $R$~(nm) & $\delta p$~(MPa) & ~$R \delta p / \gamma$~   
                            &  $R$~(nm) & $\delta p$~(MPa) & ~$R \delta p / \gamma$~ \\
        \hline
        $\Delta x = 1.0$~nm  &   5.832   &   $5.762$ &   1.023   
                             &   11.91 &   $2.804$   &   1.016  \\
        $\Delta x = 0.5$~nm  &   5.778 &   $5.728$  &   1.0075
                              &    11.89   &  $ 2.764$ &   1.0004   
                               
    \end{tabular}
    \caption{Validation of Laplace pressure measured in simulations for droplets of different radii $R$ and at two different grid resolutions; expected value for $R \delta p / \gamma$ is one.}
    \label{tab:Laplace}
\end{table}

\begin{figure}
  \centering
  \includegraphics[width=.55\columnwidth]{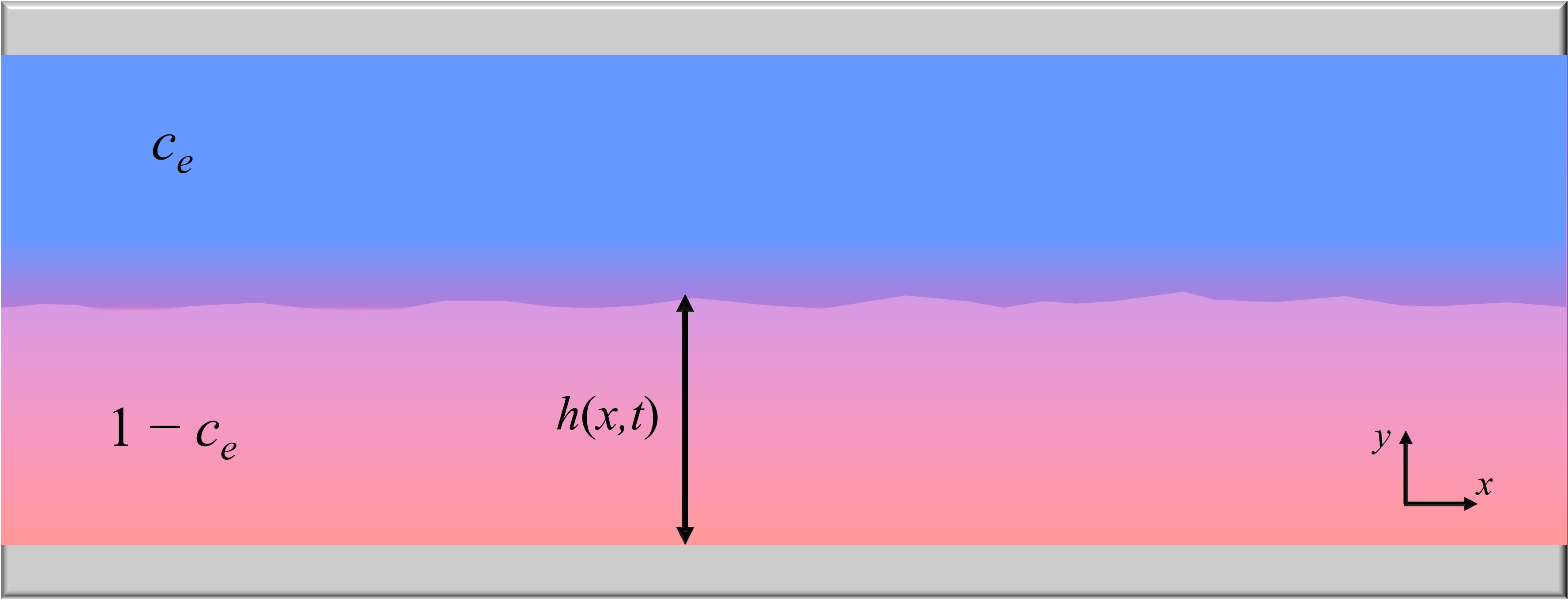}
  \caption{Illustration of system geometry for the capillary wave measurements.  Here, blue corresponds to concentration $c_e$ and red corresponds to $1-c_e$.}
  \label{fig:CapWave_Geometry}
\end{figure}

We model a physical system initialized with concentration $c_{1,e}$ in the upper half 
and $c_{2,e}$ in the lower half 
with rigid, no-slip walls at $y=0$ and $y=L_y$ (see Fig.~\ref{fig:CapWave_Geometry}).
The boundary conditions at $x=0,L_x$ are either: a) periodic; or 
b) no-flux ($\SpeciesFlux = 0$), no-slip ($u = 0$) walls with neutral wettability ($\partial_x c = 0$).
The capillary wave height is $\delta h(x,t) = h(x,t) - h_0$ where $h_0 = L_y/2$ is the unperturbed interface height.
For all of the simulations discussed here, we define $h(x,t)$ to be the lowest $y$ value at which the linear interpolant of concentration is $c(x,y,t) = 0.5$.

At thermodynamic equilibrium the statistical properties of capillary wave height are given by the equipartition of energy (see Appendix and \cite{CapWaveFluct2023}). 
In systems with periodic boundaries the variance is 
\begin{align}
    \langle \delta h(x)^2 \rangle = \frac{\ell_*^2}{12}~\frac{L_x}{L_z}
    \label{eq:Var_Periodic}
\end{align}
where $\ell_* = \sqrt{ k_B T / \gamma}$.
For systems with no flux boundary conditions the variance is
\begin{align}
    \langle \delta h(x)^2 \rangle = \ell_*^2~\frac{L_x}{L_z}
    \left[  \frac{1}{12} + \left(\frac{1}{2} - \frac{x}{L_x}  \right)^2 \right]
        \label{eq:Var_NoFlux}
\end{align}
Stochastic lubrication theory and molecular dynamics simulations are in excellent agreement with Eqns.~(\ref{eq:Var_Periodic}) and (\ref{eq:Var_NoFlux}).\cite{CapWaveFluct2023}

Similarly, the spatial correlation of capillary wave heights, $\langle \delta h(x) \delta h(x') \rangle$, in systems with periodic boundary conditions is (see Appendix)
\begin{align}
    \frac{\langle \delta h(x) \delta h(x') \rangle}{\langle \delta h(x)^2 \rangle}
    = \frac{ 
    \mathrm{Li}_2( e^{-2\pi i s/L_x}) + \mathrm{Li}_2( e^{ 2\pi i s/L_x}) 
    }{2\mathrm{Li}_2(1)}
    \label{eq:Corr_Periodic}
\end{align}
where $s = x' - x$ and $\mathrm{Li}_i(z)$ is the dilogarithm (polylogarithm function of order 2).
For system with no flux boundary conditions the correlation is
\begin{align}
    \frac{\langle \delta h(x) \delta h(x') \rangle}{\langle \delta h(x)^2 \rangle}
    = \frac{ 
    \mathrm{Li}_2( e^{-  \pi i s/L_x}) + \mathrm{Li}_2( e^{\pi i s/L_x}) 
    + \mathrm{Li}_2( e^{-\pi i (2x+s)/L_x}) + \mathrm{Li}_2( e^{\pi i (2x+s)/L_x})
    }{2 \mathrm{Li}_2(1) + 2 \mathrm{Li}_2( e^{2 \pi i x/L_x}) }
    \label{eq:Corr_NoFlux}
\end{align}
Note that these correlations are long-ranged, spanning the full width of the system, for both periodic and no-flux boundary conditions.

\begin{figure}[h!]
  \centering
  \includegraphics[width=.49\textwidth]{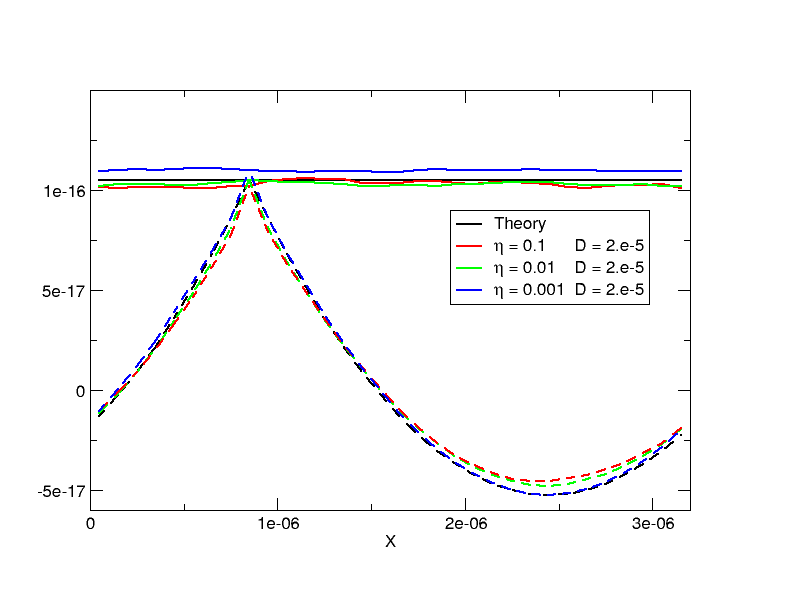} 
  \includegraphics[width=.49\textwidth]{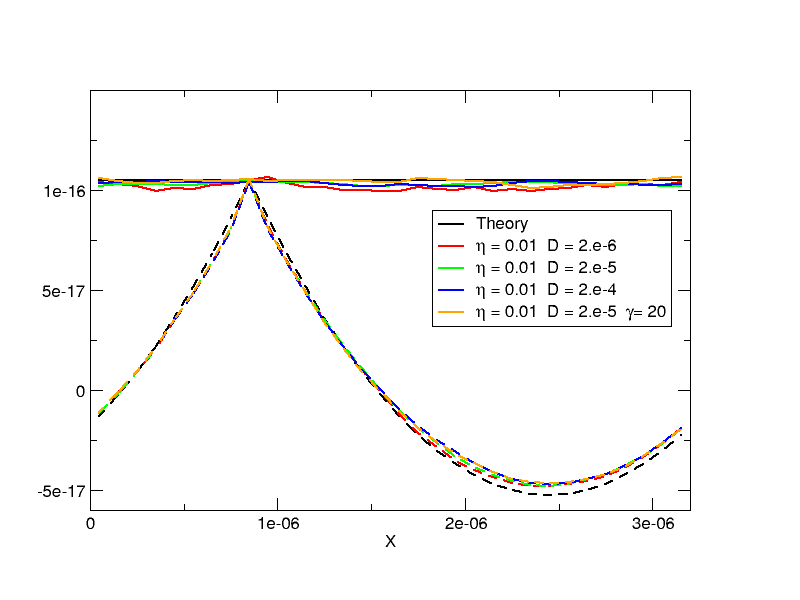}
 
  \caption{Variance (solid lines) and correlation (dashed lines) of capillary wave height for periodic boundary conditions.
  For the spatial correlations $x' = \frac14 L + \frac12 \Delta x$. 
   Left panels show results for different viscosities (see legend); right panels show results for different species diffusion and surface tension (see legend).
  Solid black lines are eqn.~(\ref{eq:Var_Periodic}) for the variances; dashed black lines are eqn.~(\ref{eq:Corr_Periodic}) for the correlations.}
  \label{fig:CapWave_VarCorrPBC}
\end{figure}

\begin{figure}[h!]
  \centering
   
 \includegraphics[width=.49\textwidth]{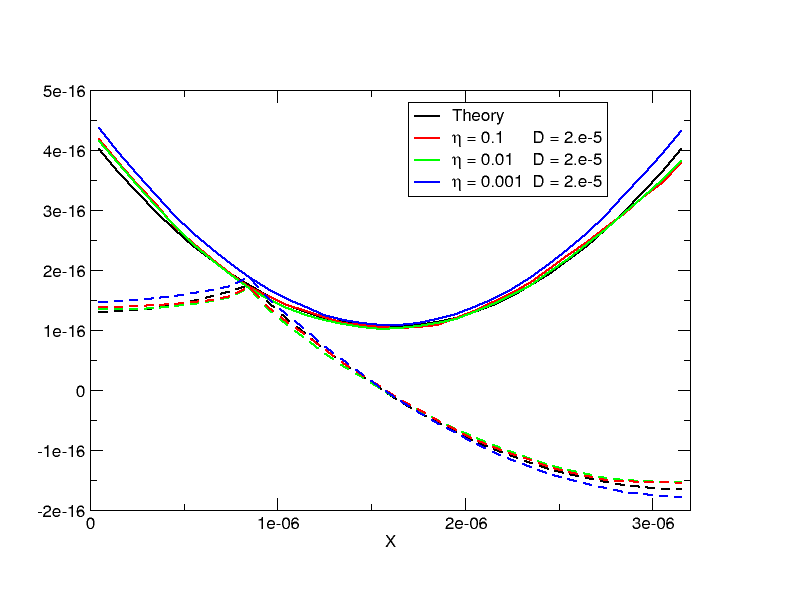} 
  \includegraphics[width=.49\textwidth]{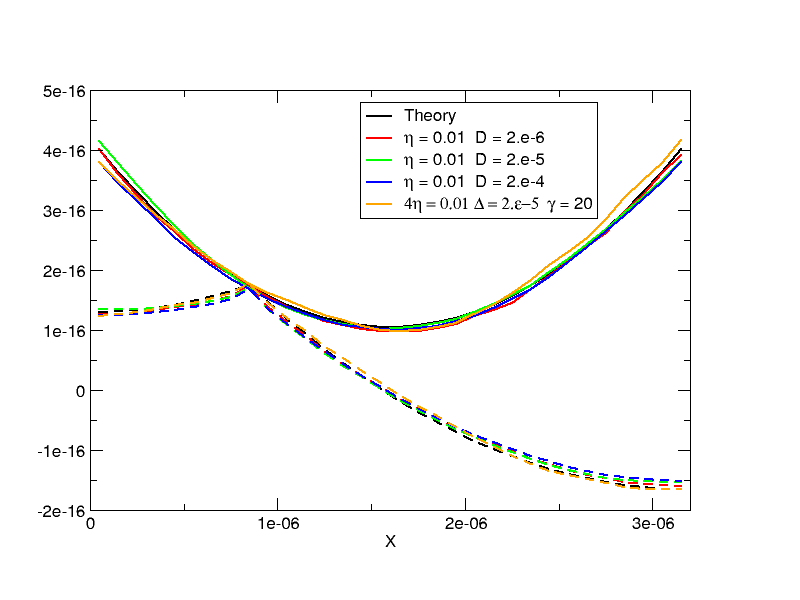}
  \caption{Variance (solid lines) and correlation (dashed lines) of capillary wave height for no flux boundary conditions.
  For the spatial correlations $x' = \frac14 L + \frac12 \Delta x$. 
   Left panels show results for different viscosities (see legend); right panels show results for different species diffusion and surface tension (see legend).
  Solid black lines are eqn.~(\ref{eq:Var_NoFlux}) for the variances; dashed black lines are eqn.~(\ref{eq:Corr_NoFlux}) for the correlations.}
  \label{fig:CapWave_VarCorrWalls}
\end{figure}

Figures~\ref{fig:CapWave_VarCorrPBC} and \ref{fig:CapWave_VarCorrWalls} show results for the variance, $ \langle \delta h(x)^2 \rangle$, and correlation, $\langle \delta h(x) \delta h(x') \rangle$, of capillary wave height measured in CH-FHD simulations using various values of viscosity. In all cases $L_x = 32$~nm and $L_y = 32$~nm. For most of these simulations, the depth of the domain was $L_z = 32$nm, corresponding to approximate 178 molecules per cell.  For the $\gamma = 20$~dyne/cm simulations we increased the mass of the molecules to $m = 3.0\times 10^{-22}$ and increased $L_z$ to 53.3333~nm, which reduces the surface tension but keeps ${\ell_*^2}/{L_x L_z}$ and the number of molecules per cell unchanged.  In all cases we see excellent agreement with capillary wave theory.

\begin{figure}[h!]
  \centering
  \includegraphics[width=.49\textwidth]{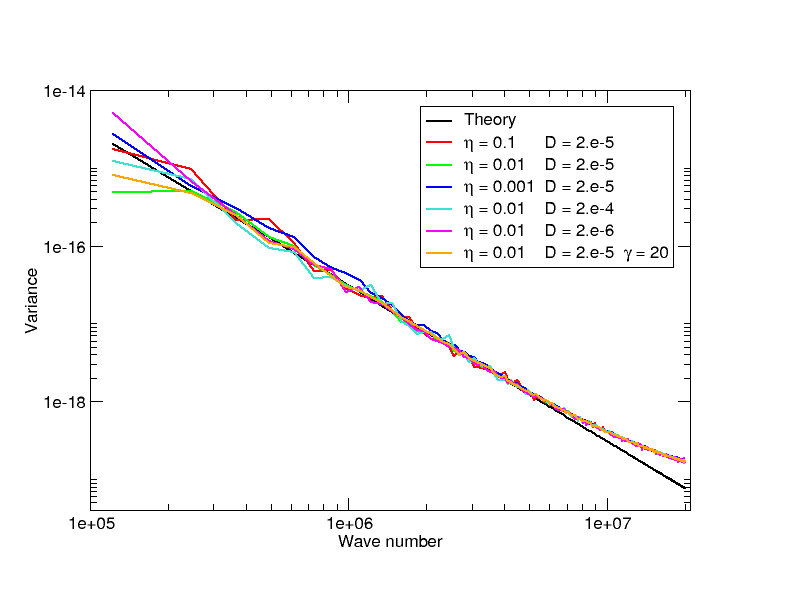}
  \includegraphics[width=.49\textwidth]{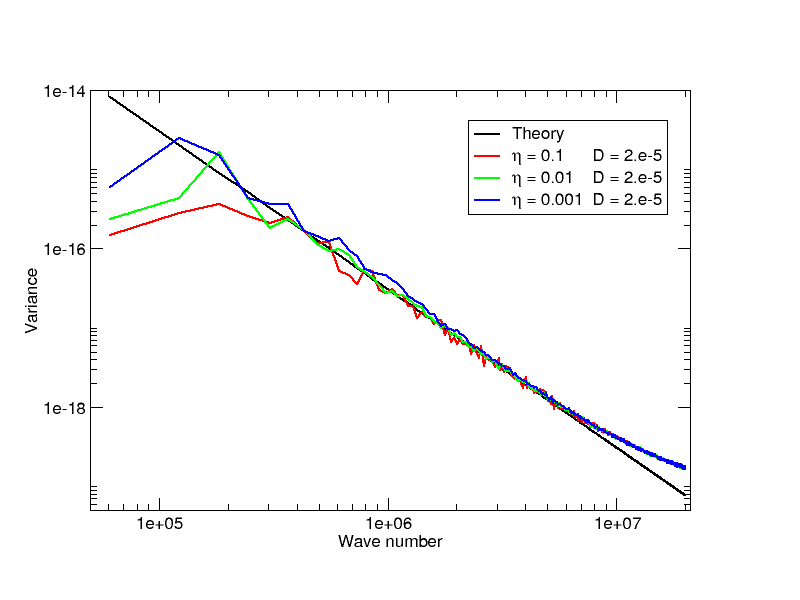}
  \caption{Spectrum of capillary wave height for: (left) periodic boundary conditions; (right) no flux boundary conditions. Black line is eqn.~(\ref{eq:Spectrum}); colored lines are from simulations (see legend). }
  \label{fig:CapWave_Spectrum}
\end{figure}

Finally, we consider the capillary wave spectrum, $\langle \delta \hat{h}(k)^2 \rangle$, where $\hat{h}(k)$ is the Fourier transform of the capillary wave height.\footnote{For no flux boundary conditions the cosine transform is used.} 
From capillary wave theory this spectrum is (see Appendix),  
\begin{align}
    \langle \delta \hat{h}(k)^2 \rangle =
    \frac{\ell_*^2}{L_x L_z}~\frac{1}{k^2} 
    \label{eq:Spectrum}
\end{align}
In all cases $L_x = 512$nm and $ L_y = 128$nm. For these simulations $L_z = 8$nm for most of the cases, corresponding to approximately 44 molecules per cell. As before, for $\gamma = 20$~dyne/cm we increased the molecular mass to $3.0 \times 10^{-22}$~g and increased the domain depth to $L_z = 13.333$~nm, which leaves the value of ${\ell_*^2}/{L_x L_z}$ unchanged.
Figure~\ref{fig:CapWave_Spectrum} shows results for various parameter values; in all cases we see excellent agreement with eqn.~(\ref{eq:Spectrum}).

\section{Concluding Remarks}

It is a truth universally acknowledged, that a statistical ensemble of states for an equilibrium system must be in want of the dynamical information required to yield transport properties.
For example, the Maxwell-Boltzmann velocity distribution tells us nothing about the thermal conductivity of a gas. 
Similarly, the equilibrium spectrum of capillary wave height must be independent of viscosity, which is not the case for the results reported by Zhang \textit{et al.}\cite{Brownian_CA_CH_PRE2024}

This discrepancy could be a numerical artefact, such as not running
the simulations for a sufficiently long time for the systems to reach
thermodynamic equilibrium.
Also, there may be an issue with their formulation of the stochastic Cahn-Hillard Navier-Stokes equations. 
A fundamental tenet of fluctuating hydrodynamics is that the amplitudes of the stochastic fluxes are obtained from their corresponding entropy production (or equivalently, the dissipation function) in order to satisfy fluctuation-dissipation relations.\cite{Zarate_07} 
It appears that this principle was not followed for the composition equation (13) in Zhang \textit{et al.}\cite{Brownian_CA_CH_PRE2024} and equation (24.4) in \cite{ElectroCH_Zhang2023} by the same authors.
If so then this could be the source of the anomalous capillary wave spectrum.

\section*{Appendix}

In this Appendix we summarize the basic derivation of the variance and correlation for the stochastic lubrication theory as discussed in \cite{CapWaveFluct2023}.
In the lubrication limit ($\delta h \ll L_x$) the elastic energy of a quasi-2D interface is
\begin{align}
    E = \frac{\gamma L_z}{2} \int_0^{L_x} \left( \frac{\partial h}{\partial x} \right)^2
    \label{eq:energy_int}
\end{align}
For periodic boundary conditions, writing the Fourier series
\begin{align}
    h(x,t) = h_0 + \sum_{n=1}^\infty a_n(t) \cos( k_n x )  + \sum_{n=1}^\infty b_n(t) \sin( k_n x )  
\end{align}
with $k_n = 2 \pi n / L_x$ gives
\begin{align}
    E = \sum_{n=1}^\infty E_n  = \pi^2 \gamma ~\frac{L_z}{L_x} \sum_{n=1}^\infty ( a_n^2 + b_n^2 ) ~n^2
\end{align}
By the equipartition of energy at equilibrium each mode with non-zero amplitude has 
an average energy of $E_n = \frac12 k_B T$ so
\begin{align}
    \langle a_n^2  \rangle + \langle b_n^2  \rangle   = \frac{1}{2\pi^2 n^2}~\frac{k_B T}{\gamma}~\frac{L_x}{L_z}
\end{align}
From this result we obtain Eqn.~(\ref{eq:Spectrum}) for the Fourier spectrum.

The variance and correlation of interface height is then given by 
\begin{align}
    \langle \delta h(x) \delta h(x') \rangle =& 
        \left\langle \left( \sum_{n=1}^\infty a_n(t) \cos( k_n x )  
        + \sum_{n=1}^\infty b_n(t) \sin( k_n x ) \right) \right. \nonumber \\
        &\times \left. \left( \sum_{m=1}^\infty a_m(t) \cos( k_m x' )  
        + \sum_{m=1}^\infty b_m(t) \sin( k_m x' ) \right) \right\rangle
\end{align}
From the definition of the polylogarithm function
\begin{align}
    \mathrm{Li}_s(z) = \sum_{k=1}^\infty \frac{z^k}{k^s}
\end{align}
and $\langle a_n a_m \rangle = \langle a_n^2 \rangle \delta_{n,m}$, $\langle b_n b_m \rangle = \langle b_n^2 \rangle \delta_{n,m}$, $\langle a_n b_m \rangle = 0$
we obtain Eqs.~(\ref{eq:Var_Periodic}) and (\ref{eq:Corr_Periodic}).

For case with walls, the Fourier series takes the form
\begin{align}
    h(x,t) = h_0 + \sum_{n=1}^\infty a_n(t) \cos( k_n x )  
\end{align}
where we now have $k_n = \pi n / L_x$.
Substituting into Eq.~(\ref{eq:energy_int} gives
\begin{align}
    E = \sum_{n=1}^\infty E_n  = \frac{\pi^2 \gamma}{4} ~\frac{L_z}{L_x} \sum_{n=1}^\infty ( a_n^2  ) ~n^2
\end{align}
Equipartition of energy at equilibrium in this again gives 
an average energy of $E_n = \frac12 k_B T$ so
\begin{align}
    \langle a_n^2  \rangle    = \frac{2}{\pi^2 n^2}~\frac{k_B T}{\gamma}~\frac{L_x}{L_z}
\end{align}
From this result we obtain Eqn.~(\ref{eq:Spectrum}) for the cosine spectrum.
Finally, we can then compute the variance and correlation from 
\begin{align}
    \langle \delta h(x) \delta h(x') \rangle = 
        \left\langle \left( \sum_{n=1}^\infty a_n(t) \cos( k_n x )  
         \right) 
         \times \left( \sum_{m=1}^\infty a_m(t) \cos( k_m x' )  
        \right) \right\rangle
\end{align}
Again, using $\langle a_n a_m \rangle = \langle a_n^2 \rangle \delta_{n,m}$, this equation can be evaluated to obtain Eqs.~(\ref{eq:Var_NoFlux}) and (\ref{eq:Corr_NoFlux}).

\section*{Acknowledgements}
The authors thank Drs. James Sprittles, Duncan Lockerby, and Jingbang Liu for fruitful discussions. 
This work was supported by the U.S. Department of Energy, Office of Science, Office of Advanced Scientific Computing Research, Applied Mathematics Program under contract No. DE-AC02-05CH11231. 

\bibliography{CapWave}

\end{document}